\documentclass[a4paper]{report}
\usepackage[utf8]{inputenc}
\usepackage[T1]{fontenc}
\usepackage{RJournal}
\usepackage{placeins}

\usepackage{graphicx}
\usepackage{amsmath,amssymb,array}
\usepackage{booktabs}
\usepackage{url}
\usepackage{lineno}
\usepackage{orcidlink}
\newcommand{\nc}{\newcommand}
\nc{\yckorcid}{\orcidlink{0000-0001-7057-2096}}
\nc{\nn}{\noindent}
\nc{\dmo}{\DeclareMathOperator}

\nc{\dc}{\definecolor}
\dc{watpink}{RGB}{198, 0, 120}
\dc{skyblue}{RGB}{0, 102, 204}
\dc{jade}{rgb}{0.0, 0.66, 0.42}
\dc{gsublue}{RGB}{0, 57, 166}
\nc{\red}[1]{\textcolor{red}{#1}}
\nc{\blue}[1]{\textcolor{blue}{#1}}
\nc{\green}[1]{\textcolor{green}{#1}}
\nc{\purple}[1]{\textcolor{purple}{#1}}
\nc{\bI}{\bm{\mathrm{I}}}
\nc{\violet}[1]{\textcolor{violet}{#1}}
\nc{\wpink}[1]{\textcolor{watpink}{#1}}
\nc{\jade}[1]{\textcolor{jade}{#1}}
\nc{\gsublue}[1]{\textcolor{gsublue}{#1}}
\nc{\watyellow}[1]{\textcolor{watyellow}{#1}}
\nc{\yck}[1]{\wpink{[YCK: #1]}}
\nc{\aimal}[1]{\gsublue{[Aimal: #1]}}

\begin{document}

\sectionhead{Contributed research article}
\volume{XX}
\volnumber{YY}
\year{20ZZ}
\month{AAAA}

\begin{article}

\fancyhf{}
\fancyhead[R]{\thepage}
\renewcommand{\headrulewidth}{0pt}
\renewcommand{\footrulewidth}{0pt}

\fancypagestyle{pagenumonly}{%
  \fancyhf{}
  \fancyhead[R]{\thepage}
  \renewcommand{\headrulewidth}{0pt}
  \renewcommand{\footrulewidth}{0pt}
}

\pagestyle{pagenumonly}

\newcommand{\ragras}{\texttt{ragR}-RAGAS}

\title{ragR: Retrieval-Augmented Generation and RAG Assessment in R}

\author{by Muhammad Aimal Rehman, Zhili Lu, Seoyoung Kim, and Chi-Kuang Yeh ~\yckorcid}

\maketitle
\thispagestyle{pagenumonly}
\pagestyle{pagenumonly}

\abstract{
Retrieval-augmented generation (RAG) combines document retrieval with large language models (LLMs) to produce responses grounded in external evidence. While several R packages support core components of RAG workflows, integrated evaluation of RAG systems in R remains limited and is often conducted through Python-based tools, most notably the RAG assessment (RAGAS) framework. To address this gap, we introduce \texttt{ragR}, an R package that unifies document ingestion, embedding and vector storage, similarity-based retrieval, grounded generation, structured question--answer logging, and RAGAS-style evaluation within a single R-native workflow. Its evaluation module, \texttt{ragR}-RAGAS, uses an LLM as a judge to score four core RAGAS metrics: context precision, context recall, faithfulness, and answer relevance. To assess the credibility of these automated judgments, we compared \texttt{ragR}-RAGAS with two independent human annotators on three document collections. The human annotators demonstrated consistently high agreement, and \texttt{ragR}-RAGAS generally aligned with their judgments, although discrepancies were more pronounced for context precision and the identification of unsupported statements. By integrating RAG construction and evaluation within a reproducible workflow in R, \texttt{ragR} provides a practical framework for research, teaching, and moderate-scale experimentation on RAG systems entirely within the R ecosystem.
}

\section{Introduction}\label{sec:intro}

\begin{figure}[ht]
\centering
\includegraphics[width=0.85\linewidth]{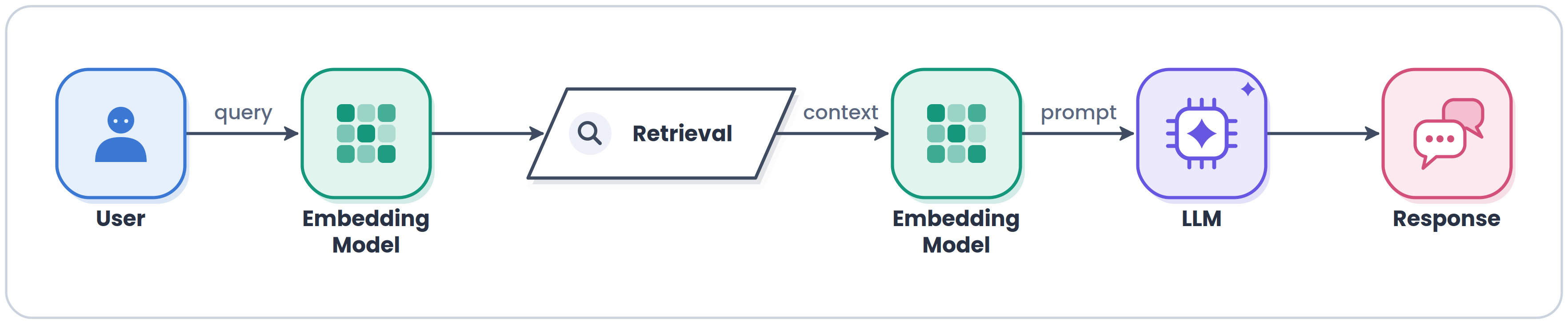}
\caption{Generic RAG pipeline: A user query is embedded, relevant context is retrieved from a vector store, and the retrieved context is added to the prompt for answer generation.}
\label{fig:rag-generic}
\end{figure}

Large language models (LLMs) have shown strong performance across a wide range of natural language tasks, but their outputs may still rely on knowledge that is outdated, incomplete, or insufficiently specialized for a given application domain \citep{lewis2020retrieval,gao2023rag}. Retrieval-augmented generation (RAG) addresses this limitation by conditioning generation on retrieved external context, which improves factual grounding and domain specificity without requiring retraining of the underlying language model. In a typical RAG pipeline, a user query is embedded, matched against a vectorized document collection, and used to retrieve relevant chunks that are then incorporated into the prompt for answer generation \citep{lewis2020retrieval}. Figure~\ref{fig:rag-generic} summarizes this workflow in three main stages: (i) query embedding, (ii) retrieval of relevant context from the vector store, and (iii) answer generation based on the original query together with the retrieved context.

Since its introduction, RAG has evolved from a relatively straightforward combination of retrieval and generation into a broader family of methods for incorporating external knowledge into language models. Research on retrieval-enhanced architectures, such as RETRO, demonstrated how large external document collections can improve language modeling without relying exclusively on larger model parameters \citep{borgeaud2022retro}. More recent approaches, including Self-RAG, have introduced adaptive retrieval and self-reflection to improve factuality and the use of retrieved evidence \citep{asai2024selfrag}. Beyond general-purpose question answering, RAG has attracted considerable interest across specialized domains, including clinical medicine, where evidence-grounded responses are particularly important \citep{zakka2024almanac}; software engineering, where repository-level context supports code completion \citep{zhang2023repocoder}; and legal applications, where precise retrieval from extensive document collections is essential \citep{pipitone2024legalbenchrag}. These developments highlight the continuing importance of RAG for knowledge-intensive applications and the corresponding need for systematic evaluation of retrieval quality, factual grounding, and answer relevance.

Despite growing interest in LLM applications within the R community, integrated support for both constructing and evaluating RAG systems remains limited in R. Recent packages such as \texttt{ragnar} \citep{ragnar} and \texttt{RAGFlowChainR} \citep{ragsflowchainr} provide useful infrastructure for building RAG workflows, including document ingestion, chunking, embedding, and retrieval over different storage backends. These tools substantially lower the barrier to developing RAG applications in R. Automated evaluation of such applications, however, is commonly performed with the RAG assessment (RAGAS) framework, whose reference implementation is written in Python \citep{es2023ragas,ragas_docs}. R users seeking automated evaluation therefore often rely on cross-language workflows.

To address this gap, we introduce \texttt{ragR}, an R package that unifies RAG construction and RAGAS-style evaluation within a single R-native workflow. The package is organized around three interoperable components: a data-ingestion module for document processing, chunking, embedding, and storage in an RDS-backed vector store; a retrieval-augmented generation module that produces grounded responses and records structured question--answer (QA) logs; and a RAGAS module that uses an LLM as a judge (LLM-as-a-judge) to score four core RAGAS metrics: context precision, context recall, faithfulness, and answer relevance. We refer to this module as \ragras{}. By linking these components through a persistent QA-log workflow, \texttt{ragR} supports reproducible experimentation and assessment entirely within the R ecosystem. Table~\ref{tab:r-package-comparison} compares \texttt{ragR} with these packages at a high level.

\begin{table}[ht]
\centering
\caption{High-level comparison of \texttt{ragR} with related R packages for RAG workflows. All three packages support RAG construction; the main distinction highlighted here is whether RAGAS-based evaluation is integrated into the same package workflow.}
\label{tab:r-package-comparison}
\small
\begin{tabular}{@{}>{\raggedright\arraybackslash}p{0.21\linewidth} >{\raggedright\arraybackslash}p{0.13\linewidth} >{\raggedright\arraybackslash}p{0.27\linewidth} >{\raggedright\arraybackslash}p{0.29\linewidth}@{}}
\toprule
\textbf{Package} & \textbf{RAGAS support} & \textbf{Storage backend} & \textbf{Workflow focus} \\
\midrule
\texttt{ragnar} & No & DuckDB-based storage; vector and BM25 retrieval & Transparent RAG construction and retrieval workflows \\
\addlinespace
\texttt{RAGFlowChainR} & No & Local and web-search oriented workflows with multiple backend options & RAG workflows with local and external retrieval backends \\
\addlinespace
\texttt{ragR} & Yes & RDS-backed vector store; collection-based retrieval & Unified RAG construction, QA logging, and RAG assessment in R \\
\bottomrule
\end{tabular}
\end{table}

Because an automated evaluator is credible only to the extent that its judgments agree with human judgment, we also assess \ragras{} against human annotation. In a validation study on three document collections, two human annotators independently judged the same assessment units as \ragras{}, and we compared their judgments at both the item level and the question level. The two annotators agreed closely with each other, and \ragras{} reproduced their judgments closely in aggregate, with weaker agreement for context precision and for identifying unsupported statements.

The remainder of the article is organized as follows. Section~\ref{sec:overview} provides an overview of the \texttt{ragR} framework. Section~\ref{sec:ingestion} describes the data-ingestion stage, and Section~\ref{sec:rag} presents the retrieval-augmented generation workflow. Section~\ref{sec:ragas} introduces \ragras{} and its metric computations. Section~\ref{sec:experiments} reports the validation study comparing \ragras{} with human annotators. Section~\ref{sec:discussion} discusses the scope and limitations of the current framework, and Section~\ref{sec:conclusion} concludes the article.

\section{Overview of the \texttt{ragR} framework}
\label{sec:overview}

The \texttt{ragR} package provides a unified R-native workflow for retrieval-augmented generation and RAGAS-style evaluation. Its main contribution is to bring document ingestion, retrieval-based answer generation, structured question--answer (QA) logging, and RAG assessment together within a single package workflow in R. This design allows users to build and study RAG applications without relying on a separate Python-based evaluation pipeline.

In this respect, \texttt{ragR} differs in scope from existing R packages for RAG. The \texttt{ragnar} package emphasizes transparent RAG construction with Markdown-based document processing, semantic chunking, embedding support from multiple providers, DuckDB-based storage, and both vector and BM25 retrieval, together with integration into \texttt{ellmer} chat workflows \citep{ragnar}. The \texttt{ragsflowchainr} package provides RAG workflows with multiple retrieval backends, including local and external database options as well as optional web search \citep{ragsflowchainr}. In contrast, \texttt{ragR} is organized around a simpler RDS-backed vector store and a persistent QA log that links grounded generation directly to downstream RAGAS-style evaluation. Its assessment module is implemented in the same spirit as the Python RAGAS framework \citep{es2023ragas,ragas_docs}, while currently focusing on four core metrics: context precision, context recall, faithfulness, and answer relevance.

\begin{figure}[ht]
\centering
\includegraphics[width=0.85\linewidth]{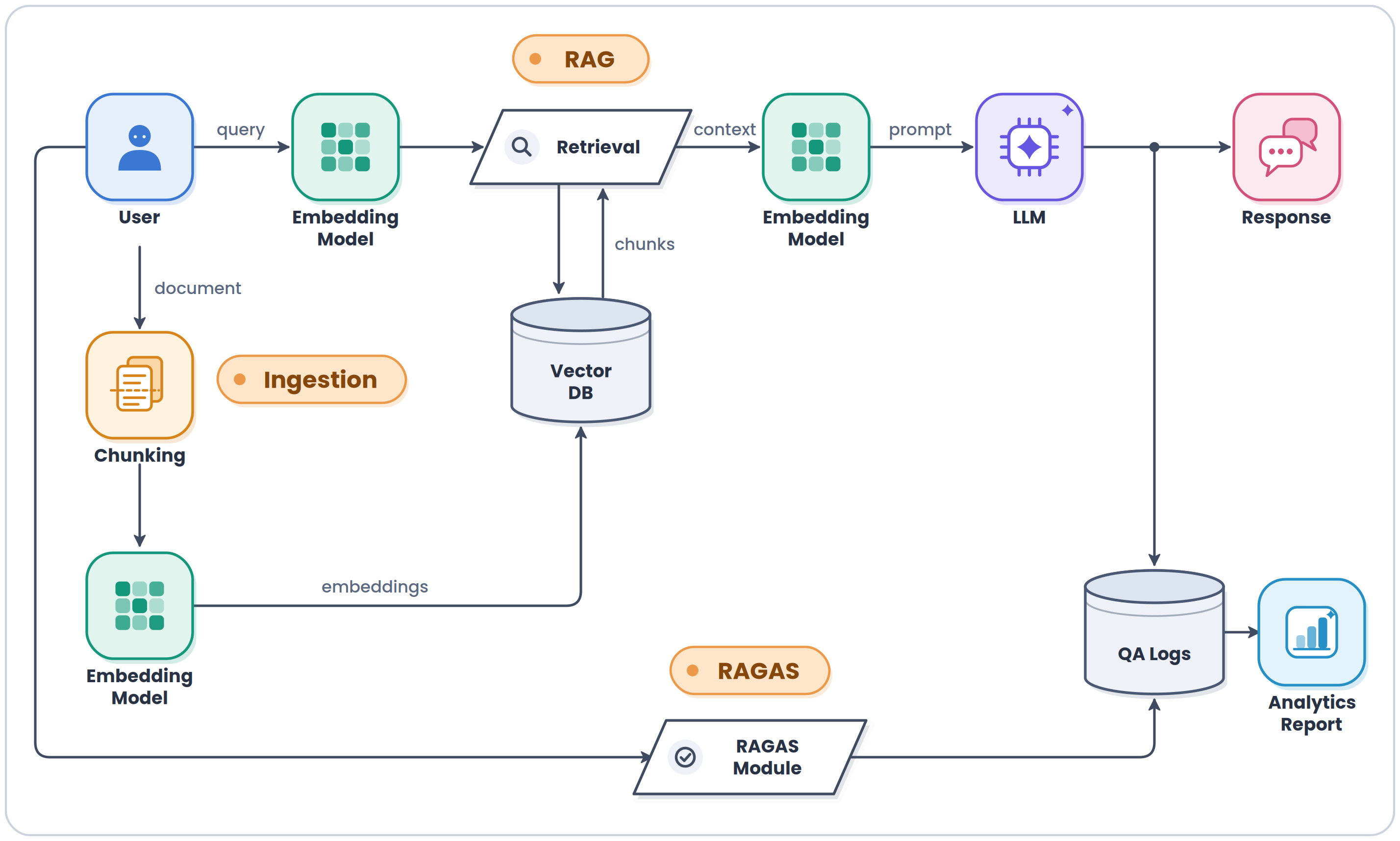}
\caption{Complete architecture of the \texttt{ragR} framework. Documents are processed during ingestion, embedded, and stored in an RDS-backed vector store. User queries are embedded and used for similarity-based retrieval over a selected collection. The retrieved context is incorporated into the prompt for language model generation. Interactions are stored in QA logs and subsequently evaluated by the RAGAS module.}
\label{fig:architecture}
\end{figure}

At a high level, the package is organized around three stages: data ingestion, retrieval-augmented generation, and RAG assessment. Figure~\ref{fig:architecture} summarizes how these stages connect within a single workflow. During ingestion, input documents are cleaned, chunked, embedded, and stored in an RDS-backed vector store. At query time, a user question is embedded and matched against a selected collection, and the retrieved context is inserted into the prompt for grounded answer generation. The resulting interaction is then written to a QA log, which serves as the input to the RAG assessment stage.

The ingestion stage prepares a document corpus for later retrieval by processing input files, segmenting the extracted text into chunks, computing embeddings, and storing the resulting records in the vector store. The current implementation supports PDF and plain-text inputs, and multiple collections can be maintained within the same RDS-backed database. The retrieval-augmented generation stage maps a user question to a grounded answer by retrieving relevant chunks from a selected collection and incorporating them into the prompt for language model generation. The generated response, together with the associated retrieval context, is recorded in a structured QA log. This logging step is central to the package design because it provides a persistent interface between answer generation and downstream evaluation. The RAG assessment stage operates on stored QA logs and computes four core RAGAS-based metrics: context precision, context recall, faithfulness, and answer relevance. In this way, \texttt{ragR} treats evaluation not as an external post-processing step, but as an integrated component of the overall workflow.

A key design feature of \texttt{ragR} is that these three stages are separate but interoperable. Ingestion is performed as preprocessing, retrieval and generation are performed at query time, and evaluation is carried out on previously recorded QA logs. This separation supports reproducible experimentation by allowing users to vary retrieval settings or prompting strategies while keeping the ingested corpus fixed, or to recompute evaluation metrics from stored interactions without rerunning the full pipeline. In the current implementation, embedding generation and chat completion are supported through OpenAI models.

The package is intended primarily for research, teaching, and moderate-scale experimentation. Its RDS-backed vector store keeps the workflow fully native to R while avoiding dependence on external vector database systems in lightweight settings. At the same time, support for multiple collections, structured QA logging, and integrated RAGAS scoring makes \texttt{ragR} suitable for end-to-end studies of how ingestion choices, retrieval settings, and prompt design affect RAG system behaviour.

\section{Data ingestion}
\label{sec:ingestion}

\begin{figure}[ht]
\centering
\includegraphics[width=0.85\linewidth]{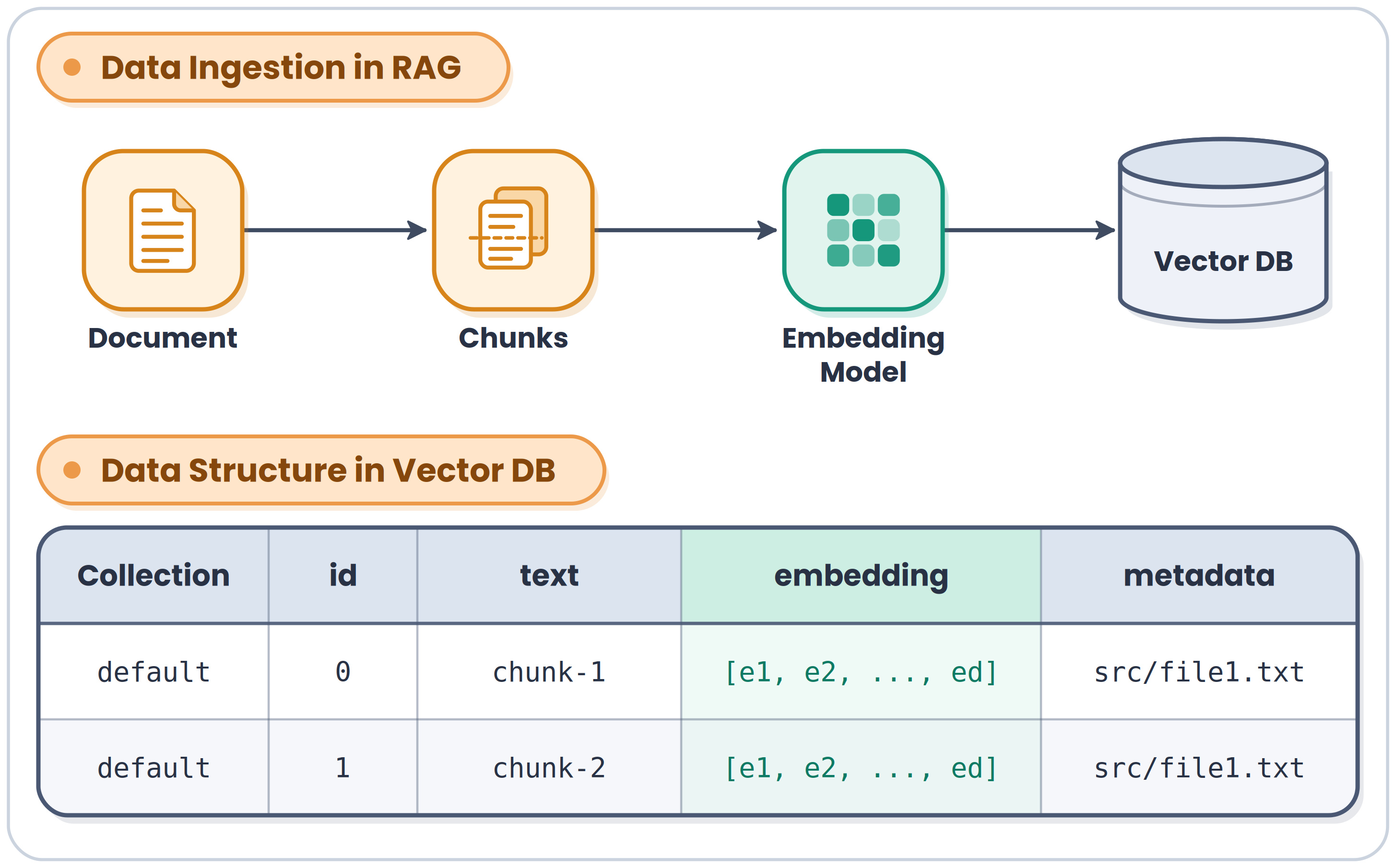}
\caption{Data ingestion workflow in \texttt{ragR} and the resulting vector store schema. Documents are chunked and embedded during ingestion and stored with identifiers and metadata for retrieval.}
\label{fig:ingestion}
\end{figure}

The ingestion stage of \texttt{ragR} prepares a document corpus for later retrieval by converting raw PDF and plain-text files into chunked text records with embeddings and metadata stored in an RDS-backed vector store. This stage is designed as a preprocessing step, so that downstream retrieval and evaluation experiments can be carried out without repeatedly reprocessing the original documents. Figure~\ref{fig:ingestion} summarizes this workflow and the resulting storage structure. In the upper part, each input document is segmented into chunks, embedded, and written to the vector store. In the lower part, the figure shows the main fields stored for each record, including the collection name, chunk identifier, text content, embedding vector, and source metadata.

The main user-facing function for this stage is \texttt{ingest\_documents()}, which carries out the ingestion workflow over one or more input files. A typical call is shown below.

\begin{verbatim}
ingestion_result <- ingest_documents(
  paths                = existing_paths,
  collection           = "default",
  chunk_size           = 1600L,
  chunk_overlap        = as.integer(round(0.30 * 1600L)),
  chunking_strategy    = "character",
  embedding_model      = "text-embedding-3-small",
  embedding_batch_size = 128L,
  embedding_max_chars  = 8000L,
  retry                = TRUE,
  max_retries          = 5L,
  resume               = TRUE,
  checkpoint_path      = NULL,
  verbose              = TRUE
)
\end{verbatim}

This interface exposes the main ingestion controls used in \texttt{ragR}. The \texttt{paths} argument specifies the input files, and \texttt{collection} determines the target collection in the vector store. The argument \texttt{chunking\_strategy} selects how documents are segmented, while \texttt{chunk\_size} and \texttt{chunk\_overlap} control the granularity of character-based chunking. The arguments \texttt{embedding\_model}, \texttt{embedding\_batch\_size}, and \texttt{embedding\_max\_chars} regulate embedding generation, and the remaining arguments support robustness through retries, resumable ingestion, and optional checkpointing.

The current implementation supports two chunking strategies. Under sentence-based chunking, each sentence is treated as a separate chunk. Under character-based chunking, documents are segmented into overlapping windows controlled by a maximum chunk length and an overlap amount. Character-based chunking is particularly useful when chunk size is treated as an experimental variable, since it provides direct control over the trade-off between fine-grained retrieval and broader contextual coverage. Overlap is used to preserve continuity across neighboring chunks by repeating boundary text, reducing the risk that related information is split too sharply at chunk boundaries.

Embedding generation is performed using the same embedding model family later used for query embeddings, ensuring that document chunks and user questions lie in a common vector space. To support efficient ingestion on moderate-sized corpora, \texttt{ragR} processes chunks in batches during embedding generation. The ingestion workflow also supports resumable execution through checkpoints, which is useful when embedding large collections or when repeated experiments are conducted over the same corpus.

The resulting vector store is persistent and remains fully native to R. Multiple collections can be stored in the same RDS-backed database, allowing users to maintain separate corpora for different domains while keeping them within a single package workflow. During retrieval, the active collection is specified explicitly, so the same underlying store can support different applications without mixing their document bases. Metadata is stored primarily for provenance and traceability, linking chunk records back to their source documents and positions.

\section{Retrieval-augmented generation}
\label{sec:rag}

\begin{figure}[ht]
\centering
\includegraphics[width=0.85\linewidth]{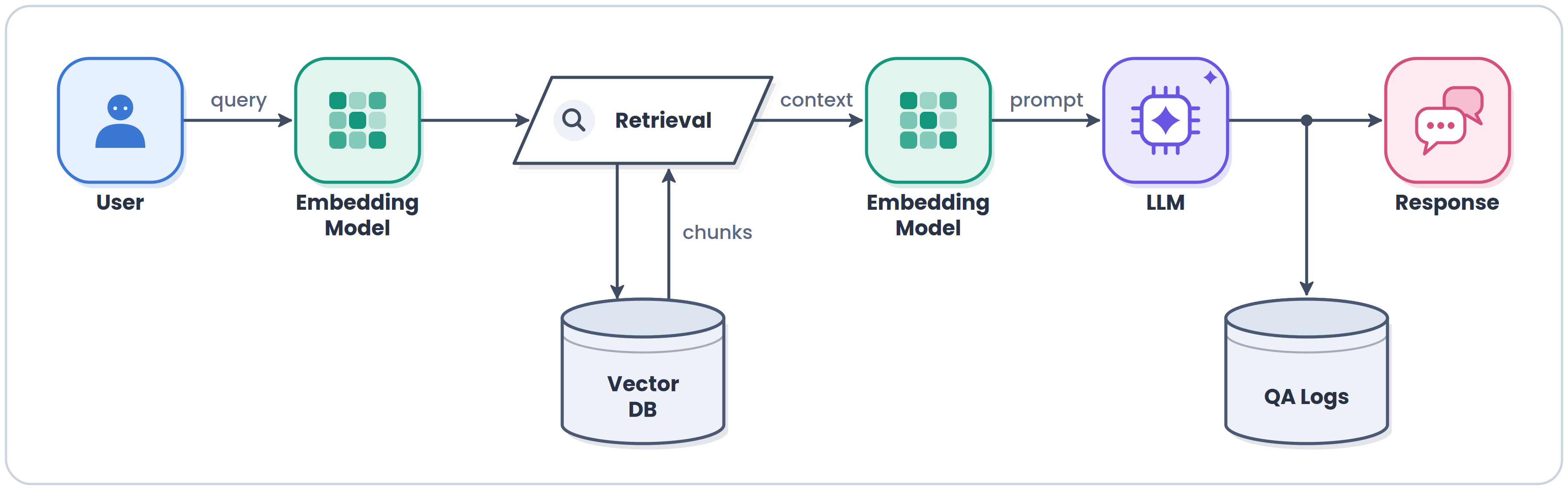}
\caption{RAG module workflow in \texttt{ragR}: query embedding, context retrieval from the vector store, prompt construction, grounded generation, and interaction logging.}
\label{fig:rag-workflow}
\end{figure}

The retrieval-augmented generation stage of \texttt{ragR} maps user questions to grounded answers by retrieving relevant document chunks from the vector store and incorporating them into the prompt for language model generation. In the package workflow, this stage is implemented primarily through the function \texttt{query\_rag()}, while structured logging of each interaction is handled separately through \texttt{log\_rag\_interaction()}. This separation keeps answer generation and evaluation-ready logging closely connected, while still allowing each stage to be inspected independently. Figure~\ref{fig:rag-workflow} summarizes this workflow. A user query is first embedded, relevant chunks are retrieved from the vector store, the retrieved context is added to the prompt for the language model, and the resulting interaction is recorded in the QA log for later evaluation.

\begin{samepage}
A typical call to \texttt{query\_rag()} is shown below.

\begin{verbatim}
res <- query_rag(
  question          = q,
  collection        = collection_name,
  top_k             = 5L,
  embedding_model   = "text-embedding-3-small",
  chat_model        = "gpt-4o-mini",
  temperature       = 0,
  max_output_tokens = 2000L,
  score_threshold   = 0,
  system_prompt     = "You are a helpful assistant."
)
\end{verbatim}
\end{samepage}

This interface exposes the main retrieval and generation controls. The \texttt{question} argument supplies the user query, and \texttt{collection} selects the document collection over which retrieval is performed. The argument \texttt{top\_k} specifies how many chunks are retrieved, while \texttt{score\_threshold} optionally filters retrieved results after similarity search. The arguments \texttt{embedding\_model}, \texttt{chat\_model}, \texttt{temperature}, and \texttt{max\_output\_tokens} control embedding generation and answer generation. The \texttt{system\_prompt} argument is optional and allows users to guide response style or behaviour.

Internally, \texttt{query\_rag()} first embeds the question using the selected embedding model, then queries the vector store over the specified collection and ranks chunks by similarity to the query embedding. The top-$k$ chunks are returned, and \texttt{score\_threshold} can optionally discard retrieved chunks whose similarity scores fall below the specified cutoff. The retrieved chunks are then passed to a prompt-construction step, and the final prompt is sent to the chat model to generate the answer. The function returns a structured result containing the generated answer, the retrieved chunks, the final prompt, and the chat model identifier. In the current implementation, retrieval is restricted by collection. The metadata stored with chunk records is preserved mainly for provenance and traceability, while collection selection provides the primary mechanism for keeping retrieval focused on a specific document set.

To support later assessment, each interaction can be appended to the QA log through \texttt{log\_rag\_interaction()}, as illustrated below.

\begin{verbatim}
qa_log <- log_rag_interaction(
  qa_log          = qa_log,
  question        = q,
  rag_result      = res,
  collection      = collection_name,
  chat_model      = chat_model,
  embedding_model = embedding_model
)
\end{verbatim}

This logging step stores the key inputs and outputs of the retrieval-augmented generation stage in a structured form suitable for downstream RAGAS evaluation. Each log entry includes a unique interaction identifier (\texttt{qa\_id}), the user question (\texttt{question}), the final prompt sent to the language model (\texttt{prompt\_final}), the model-generated answer (\texttt{answer\_model}), the reference answer when available (\texttt{answer\_reference}), the active collection (\texttt{collection}), the retrieved chunk identifiers (\texttt{retrieved\_ids}), the retrieved texts (\texttt{retrieved\_texts}), the chat and embedding model names, and a timestamp.

Because RAGAS scoring operates on stored QA logs, evaluation reflects the specific ingestion and retrieval-generation settings used to produce those interactions. If those settings change, the QA log must be regenerated before re-evaluation. Thus, each QA log represents a fixed experimental condition and provides the evaluation input for that particular RAG configuration.

In addition to programmatic use within R, \texttt{ragR} also exposes API endpoints for retrieval-augmented generation and RAG assessment through the \texttt{plumber} package \citep{plumber}. This allows the same package workflow to be accessed from lightweight external interfaces without changing the core R implementation.

\section{RAG assessment}
\label{sec:ragas}

The RAG assessment stage of \texttt{ragR} provides automated evaluation of retrieval-augmented generation systems using four core RAGAS-based metrics. Evaluation is performed on stored QA logs generated under a specified ingestion and retrieval configuration. Each log contains the question, model-generated answer, retrieved context, and reference answer. The main user-facing function for this stage is \texttt{compute\_ragas\_metrics\_llm()}, which computes metric values for each recorded QA interaction. A typical call is shown below.

\begin{verbatim}
qa_metrics <- compute_ragas_metrics_llm(
  qa_log,
  seed = 42
)
\end{verbatim}

The current implementation uses a large language model as a judge (LLM-as-a-judge) to evaluate four core metrics: context precision, context recall, faithfulness, and answer relevance. We refer to this LLM-as-a-judge implementation of RAGAS in \texttt{ragR} as \ragras{}. Section~\ref{sec:experiments} examines how closely its judgments correspond to independent human evaluation. 

\subsection{Metrics}
\label{sec:ragas-metrics}

The metrics implemented in \texttt{ragR} are intended to capture complementary aspects of retrieval quality and response quality in a retrieval-augmented generation pipeline. In particular, Context Precision and Context Recall assess the quality of the retrieved context, whereas Faithfulness and Answer Relevance assess the quality of the generated response. Considered jointly, these metrics provide a more informative assessment than any single score alone, since they help distinguish between retrieval failures, such as irrelevant or incomplete context, and generation failures, such as unsupported or unresponsive answers.

\textbf{Context Precision} evaluates whether the retrieved chunks that are useful for producing the answer are ranked near the top of the retrieved list. Each of the $K$ retrieved chunks is judged useful ($v_k = 1$) or not useful ($v_k = 0$), and the score is the average precision over retrieval rank,
\begin{equation*}
\mathrm{CP} = \frac{\sum_{k=1}^{K} v_k P_k}{\sum_{k=1}^{K} v_k}, \qquad P_k = \frac{1}{k}\sum_{j=1}^{k} v_j,
\label{eq:context-precision}
\end{equation*}
where $P_k$ is the proportion of useful chunks among the top $k$; the score is 0 when no chunk is judged useful. High values indicate that the useful chunks appear early in the ranking, whereas low values indicate that they are preceded by chunks that are not useful. Because only the ranks of useful chunks enter the calculation, chunks ranked below the last useful chunk do not affect the score, so context precision does not measure how much irrelevant material the retrieved set contains. \textbf{Context Recall}, by contrast, evaluates whether the retrieved context contains sufficient information needed to support the reference answer. High values indicate that the necessary supporting evidence is present in the retrieved context, whereas low values suggest that important information is missing. \textbf{Faithfulness} measures whether the generated answer is grounded in the retrieved context. High values indicate that the answer is supported by the retrieved evidence, whereas low values indicate unsupported claims or hallucinated content. \textbf{Answer Relevance} evaluates whether the generated answer directly addresses the user’s question. High values indicate strong alignment between the question and the answer, whereas low values reflect answers that are off-topic, vague, or incomplete. Context precision, context recall, and faithfulness lie between 0 and 1. Answer relevance is a mean cosine similarity and can therefore range from $-1$ to 1. For all four metrics, larger values indicate better performance. For comparison with human ratings, Section~\ref{sec:experimental-design} maps answer-relevance scores onto a three-point scale.

\subsection{LLM-based scoring}
\label{sec:ragas-llm}

The ragR assessment procedure follows a metric-specific sequence of prompting, structured-output parsing, and score aggregation. The metrics are not assigned through a single overall scoring prompt. Instead, LLM-based scoring operates on stored QA logs. For each interaction, the implementation uses the question, the model-generated answer, and the retrieved texts. When \texttt{answer\_reference} is available, it is used in metrics that require a reference-style target, consistent with the corresponding RAGAS workflow.

For Context Precision, retrieved chunks are evaluated individually and in retrieval order. The judge determines whether each chunk was useful in arriving at the answer, and these item-level decisions are aggregated using average precision over retrieval rank. For Context Recall, the retrieved chunks are concatenated, and the judge model determines whether each reference-answer statement is supported by the retrieved context. For Faithfulness, the answer is first decomposed into simplified factual statements, and each statement is then checked against the retrieved context to determine whether it is supported. For Answer Relevance, the judge model generates reverse questions from the answer, and the final score is computed from embedding-based similarity between the original question and the generated questions.

\section{Experiments and results}
\label{sec:experiments}

This section reports a validation study of \ragras{}. Its purpose is to establish whether the judgments produced by \ragras{} agree with independent human judgments of the same RAG outputs, not to optimise a particular RAG application. Two human annotators and \ragras{} evaluated identical assessment units drawn from three document-grounded RAG applications. We first establish the level of agreement between the two annotators and then compare \ragras{} with the annotators at two levels: individual assessment items and question-level metric scores.

\subsection{Experimental design}
\label{sec:experimental-design}

The study used three source documents that cover medical education, educational policy, and literature, and that differ in style and file format. Experiment~1 (Anatomy) used the plain-text version of \textit{Gray's Anatomy for Students} by Drake, Vogl, and Mitchell from the textbook collection of the MedQA dataset \citep{jin2021medqa}, obtained through the MedQA-USMLE dataset on Kaggle.\footnote{\url{https://www.kaggle.com/datasets/kamalesh03/medqa-usmle}} Experiment~2 (Social Media Policy) used \textit{Social Media for Learning by Means of ICT}, a policy brief by Piet Kommers published as a PDF document by the UNESCO Institute for Information Technologies in Education \citep{kommers2011}. Experiment~3 (Alice) used Lewis Carroll's \textit{Alice's Adventures in Wonderland} in the plain-text edition of Project Gutenberg \citep{carroll1865}. Table~\ref{tab:experiment-config} summarises the configuration of each experiment.

For each experiment, \texttt{ragR} ingested the source document into a separate collection, generated embeddings, and stored the resulting chunks in its vector store. Thirty questions were then submitted to each application. For every question, \texttt{ragR} retrieved the five most similar chunks and generated an answer from the retrieved material. The question, generated answer, reference answer, retrieved chunks, and associated metadata were recorded in the QA log, which \ragras{} subsequently scored. Across the three experiments, the study comprised 90 QA interactions and 450 retrieved chunks.

All questions were in scope, that is, answerable from the source document. This was a deliberate design choice, because the study evaluates the credibility of the metrics rather than the behaviour of a chatbot when faced with questions outside its knowledge base. In the experiments, the generation prompt instructed the model to reply ``I don't know the answer'' whenever a question fell outside the scope of the retrieved material. Nearest-neighbour retrieval always returns the $k$ most similar chunks, even when none of them is relevant, so the decision to abstain rests with the generator. The RAGAS metrics are not designed to score such abstentions. The answer-relevance prompt classifies ``I don't know based on the provided context'' as a noncommittal answer, assigning it a score of zero. Moreover, an abstention contains no factual statements for faithfulness to verify. Including out-of-scope questions would therefore have measured how the metrics treat refusals, rather than whether their judgments agree with human judgment.

The answer-relevance score of \ragras{} is continuous, whereas the annotators rated each answer on the three-point scale $\{0, 0.5, 1\}$. To compare the two, each \ragras{} score $s$ was mapped to the human scale as $s < 0.25 \mapsto 0$, $0.25 \le s \le 0.75 \mapsto 0.5$, and $s > 0.75 \mapsto 1$; negative scores therefore map to 0. The cut-points are the midpoints between adjacent human categories, so the mapping assigns each score to its nearest category. They were fixed before the comparison with the human ratings. All answer-relevance comparisons, at both the item and the question level, use the mapped values.

\begin{table}[htbp]
\centering
\small
\caption{Configuration of the three validation experiments. Settings that were identical across experiments span all three columns.}
\label{tab:experiment-config}
\begin{tabular}{@{}>{\raggedright\arraybackslash}p{0.31\linewidth} >{\raggedright\arraybackslash}p{0.20\linewidth} >{\raggedright\arraybackslash}p{0.20\linewidth} >{\raggedright\arraybackslash}p{0.20\linewidth}@{}}
\toprule
 & \textbf{Experiment 1} & \textbf{Experiment 2} & \textbf{Experiment 3} \\
\midrule
\multicolumn{4}{@{}l}{\textit{Corpus}} \\
Source document & \textit{Gray's Anatomy for Students}, MedQA textbook collection \citep{jin2021medqa} & \textit{Social Media for Learning by Means of ICT}, UNESCO IITE policy brief \citep{kommers2011} & \textit{Alice's Adventures in Wonderland}, Project Gutenberg eBook \#11 \citep{carroll1865} \\
Format & Plain text & PDF & Plain text \\
\addlinespace
\multicolumn{4}{@{}l}{\textit{Ingestion}} \\
Chunk size / overlap & \multicolumn{3}{l}{1{,}600 / 480 characters} \\
Embedding model & \multicolumn{3}{l}{\texttt{text-embedding-3-small}} \\
\addlinespace
\multicolumn{4}{@{}l}{\textit{Retrieval and generation}} \\
Questions & 30 & 30 & 30 \\
Retrieved chunks per question & \multicolumn{3}{l}{5} \\
Generator LLM & \multicolumn{3}{p{0.60\linewidth}}{\texttt{gpt-4o-mini} (temperature 0; at most 10{,}000 output tokens)} \\
System prompt & \multicolumn{3}{p{0.60\linewidth}}{``You are a helpful assistant. You answer fully and truthfully.''} \\
Origin of questions & \multicolumn{3}{l}{Written by the authors} \\
Origin of reference answers & \multicolumn{3}{l}{Written by the authors} \\
\addlinespace
\multicolumn{4}{@{}l}{\textit{Evaluation with ragR-RAGAS}} \\
Judge LLM & \multicolumn{3}{l}{\texttt{gpt-4o-mini} (temperature 0)} \\
Seed & \multicolumn{3}{l}{42} \\
Embedding model & \multicolumn{3}{l}{\texttt{text-embedding-3-small}} \\
Reverse questions per answer & \multicolumn{3}{l}{3} \\
Statement extraction & \multicolumn{3}{l}{By the judge LLM} \\
Context-precision verdicts & \multicolumn{3}{l}{Majority vote over 3 judge calls per chunk} \\
Evaluation runs & \multicolumn{3}{l}{3} \\
Verdict scale & \multicolumn{3}{p{0.60\linewidth}}{Agreement analysis: $\{0,1\}$ for context precision, context recall, and faithfulness; $\{0,0.5,1\}$ for answer relevance. Continuous ragR-RAGAS answer-relevance scores were discretized at cut-points 0.25 and 0.75.} \\

\addlinespace
\multicolumn{4}{@{}l}{\textit{Software}} \\
\pkg{ragR} version & \multicolumn{3}{l}{0.1.1} \\
R version & \multicolumn{3}{l}{4.5.1} \\
\bottomrule
\end{tabular}
\end{table}

\subsection{Assessment units}
\label{sec:assessment-units}

Each metric was evaluated on its own assessment unit, so the number of item-level judgments varies by metric (Table~\ref{tab:assessment-items}). For context precision, each retrieved chunk is an item, giving 150 items per experiment. For context recall, each reference-answer statement is an item, and for faithfulness, each statement extracted from a generated answer is an item; because answers and reference answers differ in length, these counts vary across experiments. Answer relevance is evaluated once for each answer, giving 30 items per experiment. The annotators judged exactly the units used by ragR-RAGAS, including the statements produced by its decomposition step. These counts are metric-specific item-level judgments, not additional QA interactions. All submitted human annotation records were validated and aligned with their corresponding experiment, question, and assessment item.

\begin{table}[htbp]
\centering
\caption{Number of assessment items in the three experiments.}
\label{tab:assessment-items}
\small
\begin{tabular}{@{}lrrrrr@{}}
\toprule
Experiment & Questions & \shortstack[r]{Context\\precision} & \shortstack[r]{Context\\recall} & Faithfulness & \shortstack[r]{Answer\\relevance} \\
\midrule
Anatomy & 30 & 150 & 62 & 131 & 30 \\
Social Media Policy & 30 & 150 & 38 & 175 & 30 \\
Alice & 30 & 150 & 30 & 77 & 30 \\
\midrule
\textbf{Total} & \textbf{90} & \textbf{450} & \textbf{130} & \textbf{383} & \textbf{90} \\
\bottomrule
\end{tabular}
\end{table}

\subsection{Human annotation}
\label{sec:human-assessment}

Two human annotators independently evaluated the same questions, generated answers, retrieved chunks, and assessment units that \ragras{} evaluated. Before beginning, both annotators received the same written annotation protocol, which defined each metric and its rating criteria. Each item was evaluated using only the information displayed in the annotation workbook. Annotators were instructed not to consult external sources, conduct web searches, use AI tools, or rely on outside domain knowledge, and not to discuss their labels with each other until both had completed their annotations. Neither annotator had access to the other annotator's ratings or to the judgments of \ragras{}. An optional Notes field was provided for genuinely ambiguous or borderline cases.

\begin{samepage}
For context precision, annotators determined whether each retrieved chunk was useful for answering the question or arriving at the reference answer. A chunk was labelled 1 (\textit{useful}) when it meaningfully helped answer the question or supported a substantive part of the reference answer, including when it supported only part of a multipart reference answer, and 0 (\textit{not useful}) otherwise. Annotators were instructed not to infer usefulness from a chunk's retrieval position.
\end{samepage}

For context recall, annotators determined whether each reference-answer statement was supported by the retrieved context. A statement was labelled 1 (\textit{supported}) when it was explicitly present in the context or followed from it by a clear and direct inference, and 0 (\textit{unsupported}) when the context provided insufficient evidence, contradicted the statement, or required outside knowledge or speculation. A compound statement containing an important unsupported claim was labelled 0 as a whole. Annotators did not judge whether the statement was the best possible answer to the question.

For faithfulness, annotators determined whether each statement extracted from the generated answer was supported by the retrieved context, labelling it 1 (\textit{supported}) or 0 (\textit{unsupported}). Paraphrase was acceptable, and exact wording was not required. Annotators judged support by the supplied context only, not general truth.

For answer relevance, annotators judged how well the complete generated answer addressed the question. A rating of 1 (\textit{fully relevant}) indicated that the answer directly addressed the question and provided the requested information. A rating of 0.5 (\textit{partially relevant}) indicated that the answer addressed only part of the question, was noticeably indirect, omitted an important component, or contained substantial irrelevant material. A rating of 0 (\textit{not relevant}) indicated that the answer failed to address the question, was essentially unrelated, or was too vague or evasive. Relevance was judged separately from factual correctness and faithfulness.

In ambiguous cases, annotators were directed to rely on the most literal evidence in the displayed material rather than filling gaps with their own knowledge. When an item remained genuinely borderline, they made their best independent judgment and could record their reasoning in the Notes field.

\subsection{Agreement measures}
\label{sec:agreement-measures}

Agreement was assessed at two levels. At the item level, we compared the two annotators with each other, which establishes whether the human reference is itself consistent, and we compared \ragras{} with each annotator. The \ragras{}--human coefficients reported below are the means of the two coefficients obtained against Annotator~1 and Annotator~2. At the question level, we compared metric scores computed from \ragras{} judgments with scores computed from the annotators' judgments.

\paragraph{Binary metrics.} For context precision, context recall, and faithfulness, each item receives a label of 0 or 1. The raw agreement $p_o$ is the proportion of items to which two raters assign the same label. Raw agreement does not account for agreement that would occur by chance, which is substantial when one label dominates. We therefore also report Cohen's kappa \citep{cohen1960} and Gwet's AC1 \citep{gwet2008}, which share the form
\begin{equation*}
\kappa = \frac{p_o - p_e^{(\kappa)}}{1 - p_e^{(\kappa)}}, \qquad
\mathrm{AC1} = \frac{p_o - p_e^{(\gamma)}}{1 - p_e^{(\gamma)}},
\label{eq:kappa-ac1}
\end{equation*}
but estimate chance agreement differently:
\begin{equation}
p_e^{(\kappa)} = \sum_{k=1}^{q} p_{1k}\,p_{2k}, \qquad
p_e^{(\gamma)} = \frac{1}{q-1}\sum_{k=1}^{q} \pi_k (1-\pi_k), \qquad
\pi_k = \frac{p_{1k} + p_{2k}}{2},
\label{eq:chance-agreement}
\end{equation}
where $p_{rk}$ is the proportion of items that rater $r$ assigns to category $k$ and $q$ is the number of categories ($q = 2$ for the binary metrics). When defined, both coefficients equal 1 under perfect agreement and 0 when observed agreement equals the estimated chance agreement.

The two coefficients use different approaches to estimating chance agreement, and both are informative here. Cohen's kappa estimates chance agreement from each rater's own label frequencies. When nearly all items receive the same label, this estimate approaches the observed agreement, and kappa is determined almost entirely by the few items in the less frequent category. Kappa can therefore be close to zero even when two raters agree on nearly every item \citep{feinstein1990}. Gwet's AC1 models chance agreement in a way that shrinks as one category becomes dominant, so it remains close to raw agreement in that situation. AC1 is less sensitive than kappa to highly imbalanced label frequencies, making the two coefficients complementary measures of agreement in this setting. Because the in-scope design produces predominantly positive judgments for context recall and faithfulness, the two coefficients are expected to diverge for these metrics. When one rater uses only one category and the other uses both, kappa equals 0. If both raters assign the same label to every item, kappa is undefined, whereas AC1 remains defined. Such cases are marked in the tables.

\paragraph{Answer relevance.} Answer-relevance ratings are ordinal, so a disagreement between 0.5 and 1 is less severe than one between 0 and 1. We report exact agreement together with the linearly weighted kappa \citep{cohen1968} and its Gwet counterpart, AC2 \citep{gwet2014}. Both use the weights $w_{jk} = 1 - |j - k|/(q - 1)$ for categories indexed $j, k = 1, \dots, q$ in increasing order, so that identical ratings receive full credit, adjacent ratings half credit, and ratings of 0 and 1 no credit. The weighted coefficients replace $p_o$ by $\sum_{j,k} w_{jk} p_{jk}$, where $p_{jk}$ is the proportion of items rated $j$ by the first rater and $k$ by the second, and replace the chance terms in Equation~\eqref{eq:chance-agreement} by $\sum_{j,k} w_{jk} p_{1j} p_{2k}$ for weighted kappa and by $\bigl(\sum_{j,k} w_{jk}\bigr)\, \sum_k \pi_k(1-\pi_k) / \{q(q-1)\}$ for AC2.

\paragraph{Question-level scores.} For each question, each annotator's item-level labels were aggregated into a question-level score using the same rule that \ragras{} applies: average precision over retrieval rank for context precision, the proportion of supported statements for context recall and faithfulness, and the rating itself for answer relevance (for \ragras{}, the mapped score). The human reference score $H_i$ for question $i$ is the mean of the two annotators' scores, and $L_i$ denotes the \ragras{} score. For each experiment and metric, we report the mean of each score over the $n = 30$ questions, the signed mean difference
\begin{equation*}
\text{Mean difference} = \frac{1}{n}\sum_{i=1}^{n} (H_i - L_i),
\label{eq:mean-difference}
\end{equation*}
so that a positive value indicates a higher human score, and the mean absolute error and root mean squared error,
\begin{equation*}
\mathrm{MAE} = \frac{1}{n}\sum_{i=1}^{n} \left| H_i - L_i \right|, \qquad
\mathrm{RMSE} = \sqrt{\frac{1}{n}\sum_{i=1}^{n} \left( H_i - L_i \right)^2 }.
\label{eq:mae-rmse}
\end{equation*}
Both are expressed on the 0--1 metric scale. MAE gives the typical size of a discrepancy, whereas RMSE gives more weight to large discrepancies. An RMSE well above the corresponding MAE therefore suggests that some questions have substantially larger discrepancies than others.

\subsection{Agreement between the human annotators}
\label{sec:human-human-agreement}

Table~\ref{tab:item-level-agreement} reports item-level agreement for all metrics and experiments. Agreement between the two annotators was consistently high. Raw agreement ranged from 95.3\% to 100\% across all experiments and metrics, and for the binary metrics Gwet's AC1 ranged from 0.918 to 0.981. For context precision, where both labels occurred frequently, Cohen's kappa was also high (0.892 to 0.934). For context recall and faithfulness, where nearly all statements were judged supported, kappa was lower (0.604 to 0.880), reflecting the small number of statements in the less frequent category rather than frequent disagreement. In Experiment~3, one annotator judged every reference-answer statement supported, so kappa for context recall is 0 by construction even though the annotators disagreed on only 1 of 30 statements.

For answer relevance, both annotators rated every answer fully relevant in Experiments~1 and~2, so their ratings agreed exactly and weighted kappa is undefined, while AC2 equals 1. In Experiment~3 the annotators disagreed on 1 of 30 answers (exact agreement 96.7\%, weighted kappa 0.792, AC2 0.982). Taken together, these results show that the annotators applied the protocol consistently across the three collections, which provides a stable human reference for evaluating \ragras{}.

\begin{table}[htbp]
\centering
\caption{Item-level agreement between the two human annotators and between \ragras{} and the annotators. Agr.\ denotes raw agreement; \ragras{}--human values are averaged across the two annotators. For answer relevance ($^a$), $\kappa$ is linearly weighted and AC1 is replaced by AC2. $^{\dagger}$ indicates $\kappa=0$ when one rater uses a single category while the other uses multiple categories; $^{\ddagger}$ indicates undefined $\kappa$ when both raters assign the same category throughout.}

\label{tab:item-level-agreement}
\small
\setlength{\tabcolsep}{5pt}
\begin{tabular}{@{}lrrrrrrr@{}}
\toprule
 & & \multicolumn{3}{c}{Human 1 vs Human 2} & \multicolumn{3}{c}{\ragras{} vs humans} \\
\cmidrule(lr){3-5}\cmidrule(l){6-8}
Metric & Items & Agr. & $\kappa$ & AC1 & Agr. & $\kappa$ & AC1 \\
\midrule
\multicolumn{8}{@{}l}{\textit{Experiment 1 (Anatomy)}} \\
Context precision & 150 & 0.967 & 0.926 & 0.939 & 0.837 & 0.664 & 0.686 \\
Context recall & 62 & 0.984 & 0.880 & 0.981 & 0.944 & 0.350 & 0.938 \\
Faithfulness & 131 & 0.954 & 0.604 & 0.948 & 0.901 & 0.257 & 0.885 \\
Answer relevance$^{a}$ & 30 & 1.000 & ---$^{\ddagger}$ & 1.000 & 0.900 & 0.000$^{\dagger}$ & 0.946 \\
\addlinespace
\multicolumn{8}{@{}l}{\textit{Experiment 2 (Social Media Policy)}} \\
Context precision & 150 & 0.973 & 0.934 & 0.955 & 0.807 & 0.594 & 0.640 \\
Context recall & 38 & 0.974 & 0.655 & 0.972 & 0.987 & 0.827 & 0.986 \\
Faithfulness & 175 & 0.954 & 0.615 & 0.948 & 0.931 & $-$0.010 & 0.926 \\
Answer relevance$^{a}$ & 30 & 1.000 & ---$^{\ddagger}$ & 1.000 & 0.800 & 0.000$^{\dagger}$ & 0.862 \\
\addlinespace
\multicolumn{8}{@{}l}{\textit{Experiment 3 (Alice)}} \\
Context precision & 150 & 0.953 & 0.892 & 0.918 & 0.877 & 0.724 & 0.777 \\
Context recall & 30 & 0.967 & 0.000$^{\dagger}$ & 0.966 & 0.817 & $-$0.029$^{\dagger}$ & 0.780 \\
Faithfulness & 77 & 0.961 & 0.749 & 0.954 & 0.903 & 0.303 & 0.887 \\
Answer relevance$^{a}$ & 30 & 0.967 & 0.792 & 0.982 & 0.817 & $-$0.066 & 0.854 \\
\bottomrule
\end{tabular}

\vspace{0.5em}
\begin{minipage}{0.95\linewidth}
\footnotesize
\end{minipage}
\end{table}

\subsection{Agreement between \ragras{} and the human annotators}
\label{sec:llm-human-agreement}

\begin{figure}[htbp]
\centering
\includegraphics[width=\linewidth]{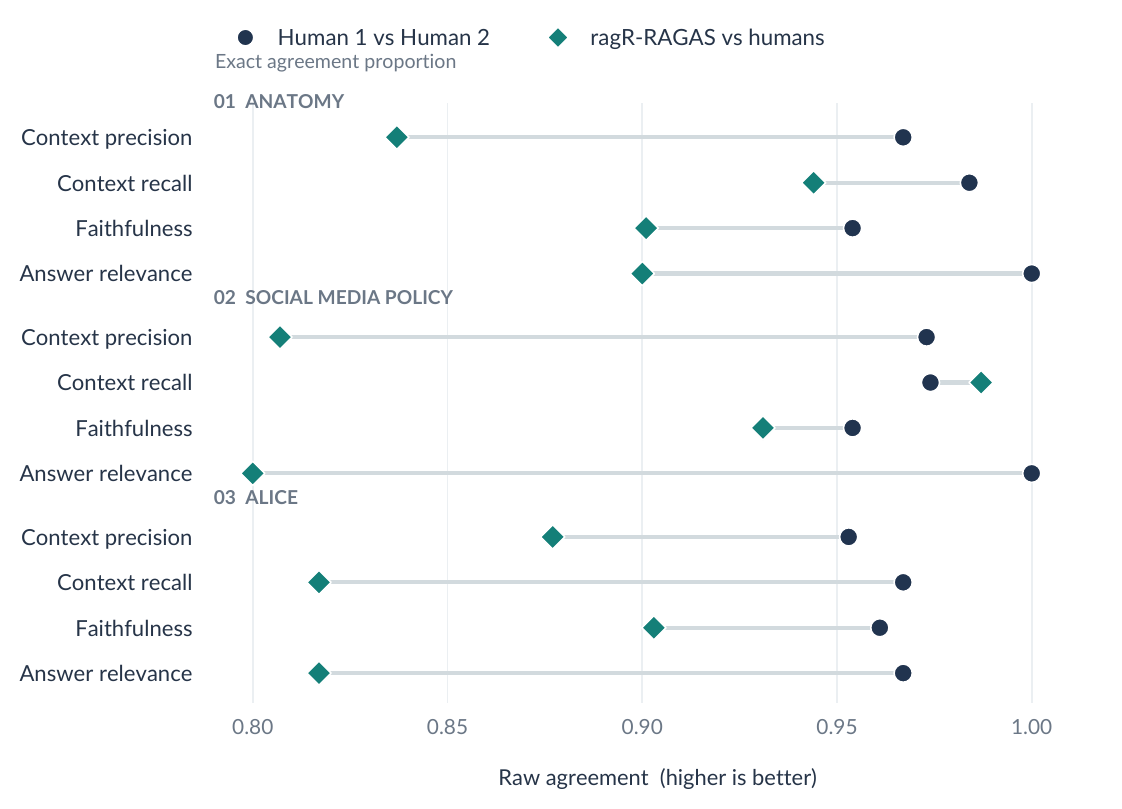}
\caption{Item-level raw agreement across three experiments. Navy circles show agreement between the two human annotators, and teal diamonds show mean agreement between \ragras{} and the annotators. Higher values indicate a larger proportion of matching judgments.}
\label{fig:item-agreement}
\end{figure}

Agreement between \ragras{} and the annotators was strong to very strong for all four metrics in all three experiments (Table~\ref{tab:item-level-agreement}). Figure~\ref{fig:item-agreement} visualizes the corresponding raw agreement values, comparing agreement between the two human annotators with mean agreement between \ragras{} and the annotators. Mean raw agreement ranged from 80.0\% to 98.7\%, and AC1 (AC2 for answer relevance) from 0.640 to 0.986. These values lie below the near-perfect agreement between the two annotators, except for context recall in Experiment~2, where \ragras{} agreed with the annotators slightly more often than the annotators agreed with each other.

For context precision, mean raw agreement ranged from 80.7\% to 87.7\% and AC1 from 0.640 to 0.777. Because both labels were common for this metric, kappa (0.594 to 0.724) closely tracked AC1, so these values are not inflated by label imbalance. For context recall, agreement was very strong in Experiments~1 and~2 (raw agreement 94.4\% and 98.7\%; AC1 0.938 and 0.986) and strong in Experiment~3 (raw agreement 81.7\%; AC1 0.780). Kappa was also high in Experiment~2 (0.827), whereas in Experiment~3 it is uninformative because one annotator judged every statement supported (Section~\ref{sec:human-human-agreement}).

For faithfulness, raw agreement (90.1\% to 93.1\%) and AC1 (0.885 to 0.926) were very strong and within 0.07 of the corresponding values between the annotators. Kappa values were lower (0.257, $-0.010$, and 0.303 for Experiments~1 to~3), which is the pattern expected when nearly all statements are supported (Section~\ref{sec:agreement-measures}): kappa then depends almost entirely on the few statements judged unsupported, on which \ragras{} and the annotators seldom coincided. For answer relevance, \ragras{} matched the annotators' ratings exactly for 80.0\% to 90.0\% of answers, with AC2 between 0.854 and 0.946, indicating strong agreement on the ordinal scale. Weighted kappa is 0 by construction in Experiments~1 and~2, where the annotators rated every answer fully relevant. In Experiment~3, it was $-0.066$, indicating that the few answers the annotators rated below fully relevant did not coincide with those rated lower by \ragras{}.

\subsection{Question-level score comparison}
\label{sec:question-level-comparison}

\begin{figure}[!htbp]
\centering
\includegraphics[width=0.75\linewidth]{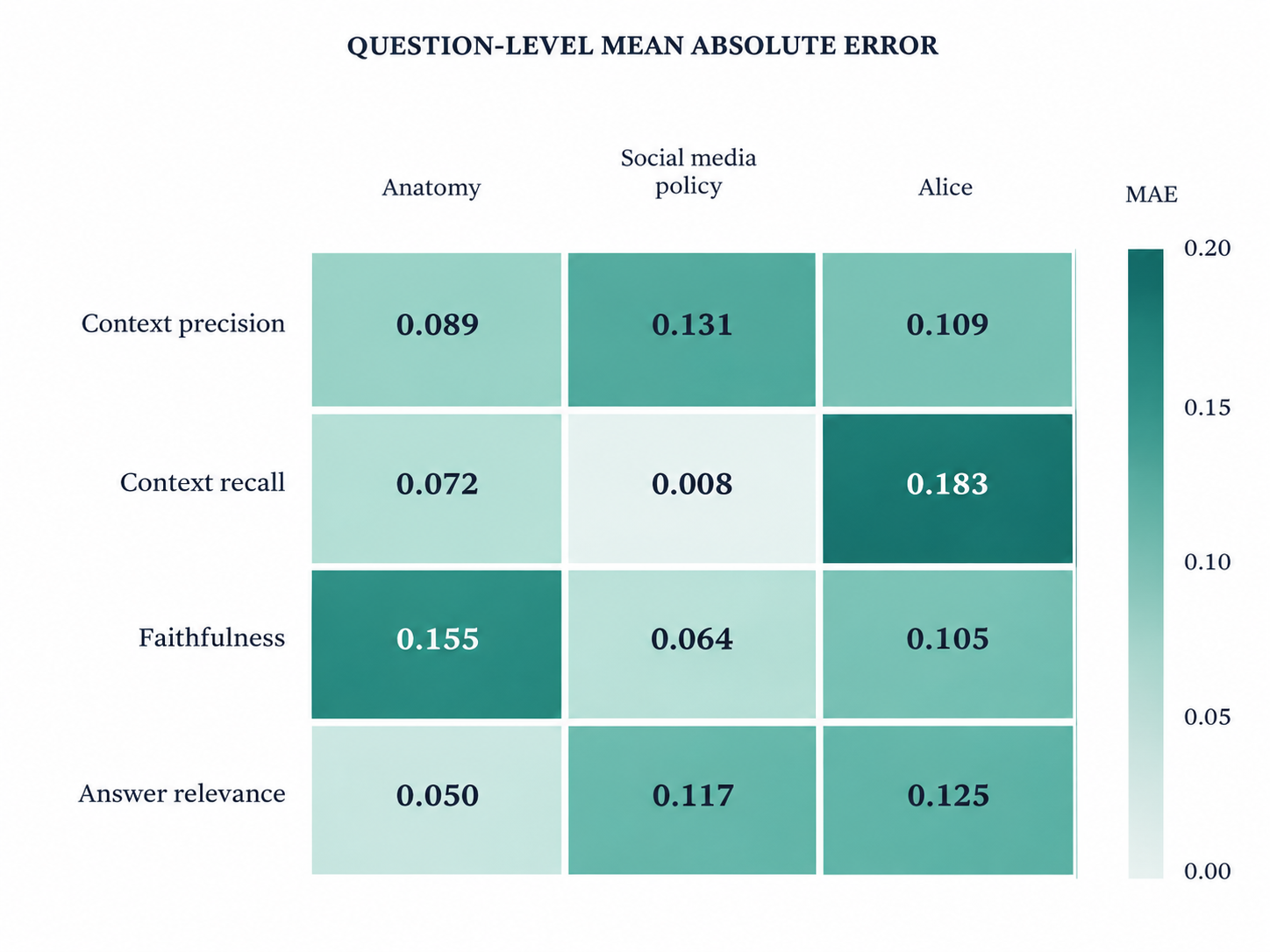}
\caption{Question-level mean absolute error (MAE) between \ragras{} scores and the mean human score across three experiments ($n=30$ questions per experiment). Darker cells indicate larger errors.}
\label{fig:question-mae}
\end{figure}

Table~\ref{tab:score-comparison} compares question-level scores. The scores assigned by \ragras{} closely matched the human scores. Figure~\ref{fig:question-mae} visualizes the corresponding mean absolute errors (MAE) across the four evaluation metrics and three experiments, highlighting where \ragras{} judgments differ most from human evaluations.
Across the twelve experiment--metric combinations, MAE ranged from 0.008 to 0.183 on the 0--1 metric scale and was at most 0.131 in ten of the twelve combinations. RMSE ranged from 0.046 to 0.418 and was at least 1.8 times MAE in every combination, suggesting that some questions had substantially larger discrepancies than others.

\begin{table}[htbp]
\centering
\caption{Question-level comparison of \ragras{} scores with the mean human score ($n = 30$ questions per experiment).}
\label{tab:score-comparison}
\small
\begin{tabular}{@{}lrrrrr@{}}
\toprule
Metric & \shortstack[r]{\ragras{}\\mean} & \shortstack[r]{Human\\mean} & \shortstack[r]{Human $-$\\\ragras{}} & MAE & RMSE \\
\midrule
\multicolumn{6}{@{}l}{\textit{Experiment 1 (Anatomy)}} \\
Context precision & 0.935 & 0.906 & $-$0.029 & 0.089 & 0.163 \\
Context recall & 0.983 & 0.911 & $-$0.072 & 0.072 & 0.226 \\
Faithfulness & 0.833 & 0.917 & 0.084 & 0.155 & 0.347 \\
Answer relevance & 0.950 & 1.000 & 0.050 & 0.050 & 0.158 \\
\addlinespace
\multicolumn{6}{@{}l}{\textit{Experiment 2 (Social Media Policy)}} \\
Context precision & 0.938 & 0.853 & $-$0.086 & 0.131 & 0.265 \\
Context recall & 0.967 & 0.958 & $-$0.008 & 0.008 & 0.046 \\
Faithfulness & 0.983 & 0.952 & $-$0.031 & 0.064 & 0.132 \\
Answer relevance & 0.883 & 1.000 & 0.117 & 0.117 & 0.274 \\
\addlinespace
\multicolumn{6}{@{}l}{\textit{Experiment 3 (Alice)}} \\
Context precision & 0.824 & 0.784 & $-$0.040 & 0.109 & 0.238 \\
Context recall & 0.833 & 0.983 & 0.150 & 0.183 & 0.418 \\
Faithfulness & 0.889 & 0.906 & 0.017 & 0.105 & 0.244 \\
Answer relevance & 0.917 & 0.958 & 0.042 & 0.125 & 0.306 \\
\bottomrule
\end{tabular}

\vspace{0.5em}
\begin{minipage}{0.95\linewidth}
\footnotesize
\end{minipage}
\end{table}

Mean scores were also close. The mean difference between the human and \ragras{} scores was within $\pm 0.09$ in ten of the twelve combinations, and context recall in Experiment~2 showed near-identical means (0.958 and 0.967, MAE 0.008). \ragras{} assigned higher mean context-precision scores than the annotators in all three experiments, with differences of $-0.029$, $-0.086$, and $-0.040$. For answer relevance, the human mean was higher in all three experiments; in Experiments~1 and~2 the annotators rated every answer fully relevant, so any rating below 1 by \ragras{} necessarily lowered its mean. For context recall and faithfulness, the direction of the differences varied across collections.



Overall, the two annotators provided a consistent reference, and \ragras{} reproduced their judgments closely. Item-level agreement with the annotators was strong to very strong for all four metrics, with AC1 and AC2 between 0.640 and 0.986, and question-level MAE did not exceed 0.183.


\FloatBarrier

\section{Discussion and limitations}
\label{sec:discussion}

In this article, we introduced \texttt{ragR}, a unified R-native framework for constructing and evaluating retrieval-augmented generation systems. By integrating document ingestion, retrieval-augmented generation, structured QA logging, and RAGAS-based evaluation within a single workflow, the package supports end-to-end experimentation in R without requiring a separate Python-based evaluation pipeline \citep{es2023ragas,ragas_docs}. We also assessed its LLM-as-a-judge module, \ragras{}, against two independent human annotators: \ragras{} reproduced the annotators' judgments closely, with strong to very strong item-level agreement for all four metrics (Section~\ref{sec:experiments}).

A central strength of \texttt{ragR} lies in its architectural unification. The QA log serves as the interface between answer generation and downstream assessment, making it possible to study how ingestion choices, retrieval settings, and prompt design affect evaluation outcomes within one consistent workflow. The package is also organized around modular stages that are separate but interoperable: ingestion is performed as preprocessing, retrieval and generation are performed at query time, and evaluation is performed on stored QA logs. This structure supports reproducible experimentation because later-stage settings can be varied while earlier stages are held fixed. More broadly, the fully R-native design lowers the barrier for users who wish to prototype and study RAG workflows directly within the R ecosystem.

Another strength of \texttt{ragR} is that the workflow remains native to R while still supporting persistent storage and reuse of intermediate artifacts. The RDS-backed vector store provides storage for embeddings and associated metadata without requiring an external vector database, and support for multiple collections allows users to maintain separate document sets for different domains within a single package workflow. These design choices make the package particularly suitable for research, teaching, and moderate-scale experimentation.

At the same time, the current framework has several limitations. First, the current vector-store design is intended for lightweight and moderate-scale use rather than large-scale deployment. For very large corpora or latency-sensitive applications, more specialized external database systems may be preferable. Second, the current implementation supports PDF and plain-text inputs, but not yet a broader range of document formats such as word-processing files. Third, the current retrieval workflow uses collection selection as its main document-level restriction, while richer metadata-aware filtering during retrieval has not yet been integrated. Fourth, the retrieval-augmented generation workflow is interaction-based rather than conversational: each logged interaction consists of a single question and a single answer, without a persistent conversational buffer across multiple user turns. Finally, the current RAG assessment module focuses on four core RAGAS metrics. While these provide a useful foundation for evaluating retrieval quality and answer grounding, they do not cover the full range of assessment possibilities in the broader and evolving RAGAS ecosystem \citep{es2023ragas,ragas_docs}.

The validation study in Section~\ref{sec:experiments} also has a defined scope. All questions were answerable from the source documents, so the study characterizes \ragras{} on RAG applications that largely worked as intended: most statements were judged supported and most answers relevant. This design establishes that \ragras{} agrees closely with human judgment under the in-scope conditions examined in this study. Notably, the strong agreement with independent human annotators was achieved using \texttt{gpt-4o-mini}, demonstrating that effective RAG assessment does not necessarily require a larger LLM. Although using the same model for generation and evaluation may introduce model-specific tendencies, the independent human comparisons provide external validation of the resulting judgments. Evaluating alternative LLM backbones presents an opportunity for future work, where more capable models may offer further improvements, while smaller models may reveal useful trade-offs between evaluation accuracy and computational cost.

These considerations also help clarify the role of \texttt{ragR} relative to existing tools. Within the broader ecosystem, Python remains the dominant environment for RAG experimentation and evaluation, and the reference RAGAS implementation is Python-based \citep{es2023ragas,ragas_docs}. Existing R packages such as \texttt{ragnar} and \texttt{RAGFlowChainR} provide useful infrastructure for retrieval-augmented generation workflows \citep{ragnar,ragsflowchainr}, but their main emphasis is on document ingestion, retrieval, and answer generation rather than integrated RAG assessment. The contribution of \texttt{ragR} lies in unifying RAG pipeline construction and RAGAS-based evaluation within a single reproducible R workflow, and in validating its LLM-as-a-judge metrics against independent human annotation.

Future development could strengthen the package in several directions, including metadata-aware filtering during retrieval, support for local language models, extension of the workflow to multi-turn conversational settings, broader implementation of RAGAS metrics described in \citet{es2023ragas} and the RAGAS documentation \citep{ragas_docs}, and support for additional input formats such as Word documents.

Overall, \texttt{ragR} is best viewed as a research and teaching framework for studying retrieval-augmented generation systems within R. Its main contribution is not large-scale deployment infrastructure, but rather the integration of RAG pipeline construction and RAGAS-based evaluation within a single, reproducible R workflow.


\section{Conclusion}
\label{sec:conclusion}

This article introduced \texttt{ragR}, an R package that unifies document ingestion, grounded generation, persistent QA logging, and RAGAS-based evaluation within a single R-native workflow. Because generation and evaluation are linked through stored QA logs, the package supports reproducible comparisons of design choices such as chunking strategies, retrieval settings, and prompting configurations.

A validation study with two independent human annotators on three document collections showed that \ragras{} reproduces human judgments closely, with strong to very strong item-level agreement for all four metrics and question-level mean absolute errors between 0.008 and 0.183. These results support the use of \ragras{} for assessment and experimentation. \texttt{ragR} thus offers R users a coherent and accessible platform for research, teaching, and reproducible experimentation on retrieval-augmented generation and its evaluation.

\bibliography{ragR}

\paragraph{Package source code.} The \texttt{ragR} package is available from the Comprehensive R Archive Network (CRAN) at \url{https://CRAN.R-project.org/package=ragR}.

\noindent\textbf{Supplementary materials.}
The experimental data, human annotation records, and statistical analysis scripts for Experiment 1 (Anatomy) are available on Zenodo at \url{https://doi.org/10.5281/zenodo.24087906}.


\address{Muhammad Aimal Rehman\\
  Department of Mathematics and Statistics\\
  Georgia State University\\
  25 Park Place, Atlanta, GA 30303\\
  United States of America\\
  \email{mrehman3@student.gsu.edu}}

\address{Zhili Lu\\
  Department of Mathematics and Statistics\\
  Georgia State University\\
  25 Park Place, Atlanta, GA 30303\\
  United States of America\\
  \email{zlu9@gsu.edu}}

\address{Seoyoung Kim\\
  Department of Mathematics and Statistics\\
  Georgia State University\\
  25 Park Place, Atlanta, GA 30303\\
  United States of America\\
  \email{skim375@student.gsu.edu}}

\address{Chi-Kuang Yeh\\
  Department of Mathematics and Statistics\\
  Georgia State University\\
  25 Park Place, Atlanta, GA 30303\\
  United States of America\\
  ORCID: \href{https://orcid.org/0000-0001-7057-2096}{0000-0001-7057-2096}\\
  \email{cyeh@gsu.edu}}

\thispagestyle{pagenumonly}
\end{article}

\end{document}